\PassOptionsToPackage{prologue,dvipsnames}{xcolor}
\documentclass[sigconf]{acmart}
\AtBeginDocument{%
  \providecommand\BibTeX{{%
    \normalfont B\kern-0.5em{\scshape i\kern-0.25em b}\kern-0.8em\TeX}}}

\usepackage{multirow}
\usepackage{tcolorbox}
\tcbuselibrary{listings}

\copyrightyear{2024} 
\acmYear{2024}
\setcopyright{acmlicensed}\acmConference[FORGE '24]{AI Foundation Models
and Software Engineering}{April 14, 2024}{Lisbon, Portugal}
\acmBooktitle{AI Foundation Models and Software Engineering (FORGE '24),
April 14, 2024, Lisbon, Portugal}
\acmDOI{10.1145/3650105.3652295}
\acmISBN{979-8-4007-0609-7/24/04}

\begin{document}

\title{On Evaluating the Efficiency of Source Code Generated by LLMs}


\author{Changan Niu}
\affiliation{
  \institution{State Key Laboratory for Novel Software Technology\\Nanjing University}
  \city{Nanjing}
  \country{China}}
\email{niu.ca@outlook.com}

\author{Ting Zhang}
\affiliation{
  \institution{School of Computing and Information Systems\\Singapore Management University}
  \country{Singapore}}
\email{tingzhang.2019@phdcs.smu.edu.sg}

\author{Chuanyi Li}
\affiliation{
  \institution{State Key Laboratory for Novel Software Technology\\Nanjing University}
  \city{Nanjing}
  \country{China}}
\email{lcy@nju.edu.cn}

\author{Bin Luo}
\affiliation{
  \institution{State Key Laboratory for Novel Software Technology\\Nanjing University}
  \city{Nanjing}
  \country{China}}
\email{luobin@nju.edu.cn}

\author{Vincent Ng}
\affiliation{
  \institution{Human Language Technology Research Institute\\University of Texas at Dallas}
  \city{Richardson}
  \state{Texas}
  \country{USA}}
\email{vince@hlt.utdallas.edu}

\begin{abstract}

Recent years have seen the remarkable capabilities of large language models (LLMs) for code generation. Different from existing work that evaluate the correctness of the code generated by LLMs, we propose to further evaluate its efficiency. More efficient code can lead to higher performance and execution efficiency of programs and software completed by LLM-assisted programming. First, we evaluate the efficiency of the code generated by LLMs on two benchmarks, HumanEval and MBPP. Then, we choose a set of programming problems from the online judge platform LeetCode to conduct a more difficult evaluation. Finally, we explore several prompts that would enable LLMs to generate more efficient code.
\end{abstract}





\maketitle

\section{Introduction}
\label{section:introduction}

With the advent of large language models (LLMs) and abundant source code data, program synthesis has entered a new era, aiming to automatically generate correct and compliant code from natural language descriptions. Extensive work has demonstrated the ability of LLMs to excel at generating well-compliant code~\cite{huang2023anpl,zelikman2023parsel,chen2023codet}. OpenAI's GPT-4 achieves 67.0\% Pass@1 on HumanEval~\cite{openai2023gpt4}, a key benchmark for measuring functional correctness by given the natural language description~\cite{chen2021humaneval}. Other open source LLMs like Code Llama~\cite{rozière2023codellama} and WizardCoder~\cite{luo2023wizardcoder} also demonstrate impressive results, with Code Llama reaching up to 53\% and 55\% on HumanEval and MBPP~\cite{austin2021mbpp}, and WizardCoder achieving 57.3\% and 51.8\% on the same benchmarks with just 7B parameters.

Given LLM's impressive performance in code generation, a number of LLM-based programming assistance tools have emerged, such as GitHub Copilot~\cite{copilot}, and JetBrains' AI Assistant~\cite{jetbrains}. These tools can offer intelligent code suggestions that automatically complete the code based on the context and the programmer's intent, thus making coding faster and speeding up the development process.

\begin{figure}
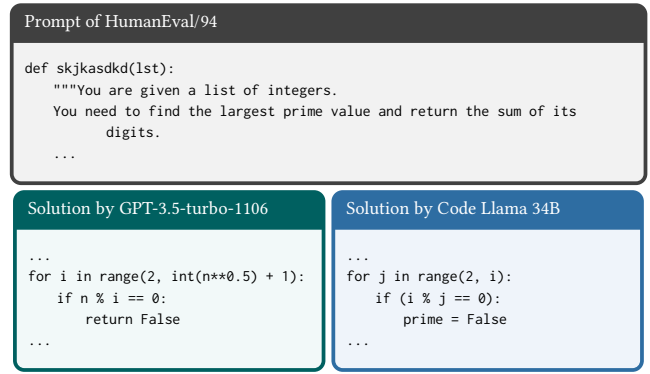

    \centering

\begin{tcblisting}{
    title={Prompt of HumanEval/94}, 
    fonttitle=\footnotesize,
    listing only,
    left=0.5mm, right=0.5mm,top=0mm,bottom=0mm,
    listing options={
        basicstyle=\scriptsize\ttfamily,
        breaklines=true,
        columns=fullflexible
    }
}
def skjkasdkd(lst):
    """You are given a list of integers.
    You need to find the largest prime value and return the sum of its digits.
    ...
\end{tcblisting}
\vspace{-0.5em}
\tcbset{nobeforeafter}
\begin{tcblisting}{
    title={Solution by GPT-3.5-turbo-1106},
    fonttitle=\footnotesize,
    listing only,
    left=0.5mm, right=0.5mm,top=0mm,bottom=0mm,
    width=0.49\linewidth,
    colback=teal!5!white,colframe=teal!75!black,
    listing options={
        basicstyle=\scriptsize\ttfamily,
        breaklines=true,
        columns=fullflexible,
    }
}
...
for i in range(2, int(n**0.5) + 1):
    if n % i == 0:
        return False
...
\end{tcblisting}
~
\begin{tcblisting}{
    title={Solution by Code Llama 34B},
    fonttitle=\footnotesize,
    listing only,
    left=0.5mm, right=0.5mm,top=0mm,bottom=0mm,
    width=0.49\linewidth,
    colback=RoyalBlue!5!white,colframe=RoyalBlue!75!black,
    listing options={
        basicstyle=\scriptsize\ttfamily,
        breaklines=true,
        columns=fullflexible
    }
}
...
for j in range(2, i):
    if (i % j == 0):
        prime = False
...
\end{tcblisting}

    \vspace{-1em}
    \caption{Code snippets extracted from the LLM-generated code for HumanEval.}
    \label{figure:example}
    \vspace{-1.5em}
\end{figure}

However, the efficiency of the generated code is overlooked. In Figure~\ref{figure:example}, GPT-3.5 and Code Llama's solutions on the HumanEval/94 example both yield correct code. However, GPT-3.5's solution exhibits higher running efficiency due to its $O(\sqrt{n})$ complexity compared to Code Llama's $O(n)$ complexity for determining prime numbers. This highlights the potential differences in execution efficiency among LLM-generated code. Recommending more efficient code not only enhances program/software performance but also increase the probability of code acceptance by developers, reducing the need for further optimization and boosting development productivity. Therefore, investigating and discussing the efficiency of LLM-generated code is essential, assuming the functional correctness of the code is ensured.

Consequently, in this paper, we propose to conduct an empirical study on the efficiency of LLM-generated code by investigating the following research questions (RQs):

\noindent \textbf{RQ1}: How efficient is the code generated by LLMs?

\noindent \textbf{RQ2}: How to prompt LLMs for more efficient code?

For RQ1, we measure and compare the execution time (we abbreviate this to ``runtime'' in this paper) of the code generated by LLMs first on two entry-level programming benchmarks, HumanEval and MBPP and then on a benchmark containing more complex problems. For RQ2, we try various prompts to explore how to make LLM generate code that executes more efficiently. Results show that simple prompts enhance efficiency for basic problems, while complex problems benefit from a chain-of-thought prompt.

This paper makes three contributions: (1) evaluate the efficiency of the code generated by LLMs. The results may guide practitioners in choosing the most suitable model based on their specific requirements,
(2) propose a LeetCode-based benchmark which provides a reference point for comparing the correctness and efficiency of more complex code, 
(3) investigate to prompt LLM for generating more efficient code, which could directly benefit developers and organizations using these models in various applications. We also make code, data and other artifacts available online~\cite{efficiencyeval}.

\section{Approach and Experiments}
\label{section:exp}

In this section, we describe how we design and conduct experiments to investigate and answer two RQs.

\subsection{RQ1: Efficiency of LLM-generated Code}
\label{section:exp_rq1}

\subsubsection{Datasets}
\label{section:exp_rq1_datasets}

We evaluate the efficiency of LLM-generated code using two entry-level programming benchmarks, HumanEval and MBPP, and a benchmark containing more complex problems.

\textit{HumanEval and MBPP}. HumanEval is used to measure functional correctness for synthesizing programs from docstrings. It consists of 164 original programming problems in Python, assessing language comprehension, algorithms, and simple mathematics, with some comparable to simple software interview questions. MBPP consists of a set of crowd-sourced Python programming problems, designed to be solvable by entry-level programmers, covering programming fundamentals, standard library functionality, etc.

\textit{LeetCodeEval}. 
LeetCode~\cite{leetcode} is a popular online judge platform that offers a wide range of problems. For each problem, LeetCode has a huge number of test cases covering a whole range of input sizes and scenarios. For accepted code, LeetCode will also give its runtime and the percentage of total code that it beats. Therefore, we propose to use LeetCode problems and the LeetCode platform to evaluate the correctness and efficiency of LLM-generated code.

In order to avoid data leakage, we select only problems from May 2023 and later (this is the latest GPT-4 knowledge cut-off). Besides, we filter out problems with images in the description and those have more downvotes than upvotes. Then, we divide the problems into three subsets according to difficulty levels officially given by LeetCode: easy, medium and hard. For each problem, we collect its URL, title, description, examples, constraints, code templates, etc. Ultimately, we build LeetCodeEval, a dataset for evaluating code correctness and efficiency based on the LeetCode platform, which consists of 44, 85 and 33 easy, medium and hard problems.

\begin{table}[t!]
    \centering
    \caption{Dataset statistics. The prompt length is the GPT-2 tokenizer length.}
    \label{table:dataset_statistics}
    \resizebox{\linewidth}{!}{%
        \begin{tabular}{lrrrrr}
            \toprule
                \multirow{2}{*}{Item} &
                \multicolumn{1}{c}{\multirow{2}{*}{HumanEval}} &
                \multicolumn{1}{c}{\multirow{2}{*}{MBPP}} &
                \multicolumn{3}{c}{LeetCodeEval} \\
            \cline{4-6}
                &
                \multicolumn{1}{c}{} &
                \multicolumn{1}{c}{} &
                \multicolumn{1}{c}{Easy} &
                \multicolumn{1}{c}{Medium} &
                \multicolumn{1}{c}{Hard} \\
            \midrule
                \# of Problems          & 164                        & 399                   & 48        & 85        & 33       \\
            \hline
                Mean Prompt Length      & 170.33                     & 52.07                 & 540.27    & 623.04    & 720.00   \\
                Median Prompt Length    & 145.5                      & 47                    & 530       & 583       & 665      \\
            \hline
                Mean \# of Test Cases   & 9.57                       & 3.10                  & 1840.10   & 1286.2    & 1036.64   \\
                Median \# of Test Cases & 7                          & 3                     & 906       & 785       & 774       \\
            \bottomrule
        \end{tabular}%
    }
\end{table}

The statistics of datasets are presented in Table~\ref{table:dataset_statistics}.

\subsubsection{Study Design}
\label{section:exp_rq1_design}

\textit{HumanEval and MBPP}. First, we use the source code provided by Liu et al.~\cite{liu2023evalplus} to let the LLM generate responses for each problem by inputting the unfinished code prompt. Since only correct code can be used in our comparison, we make the LLM generate $k$ responses for each problem to improve the possibility of collecting correct code. Then, we execute each code on corresponding test cases to determine if it is correct. If any one of the $k$ codes generated by a LLM passes all the test cases, we consider the LLM to have passed the problem, and take the first passing code for efficiency evaluation in the next step. Otherwise, the LLM is considered to have failed on that problem. Next, we measure the runtime of each selected correct code. However, executing the code on real hardware directly will introduce a lot of noise due to machine loads, configurations, etc.~\cite{madaan2023learning}. Therefore, we utilize the gem5 CPU simulator~\cite{binkert2011gem5}, which is the mostly used golden standard in both academia and industry and is able to ensure the evaluation progress reliable and reproducible~\cite{akram2019validation,madaan2023learning}. To ensure the evaluation statistically significant, we repeat the execution of each piece of code for 10 times and take the average runtime as the final result.

\begin{figure}
    \centering
\begin{tcolorbox}[fontupper=\footnotesize\ttfamily,left=1mm,right=1mm,top=1mm,bottom=1mm]
Please solve the following programming problem entitled ``\{title\}'' in C++, the problem is described below:\\
\\
\{description\}\\
\\
\{examples\}\\
\\
\{constraints\}\\
\\
Please use the following code template:\\
\textasciigrave\textasciigrave\textasciigrave cpp\\
\{code\_template\}\\
\textasciigrave\textasciigrave\textasciigrave
\end{tcolorbox}
\vspace{-1.5em}
    \caption{The prompt template for LeetCodeEval.}
    \label{figure:leetcode_prompt}
\end{figure}

\textit{LeetCodeEval}. We focus on one of the performance-oriented languages, i.e., C++. For each problem in LeetCodeEval, we first prompt LLMs to generate 3 different C++ codes. Figure~\ref{figure:leetcode_prompt} present the prompt template obtained by asking ChatGPT. For each generated code, we submit it to the LeetCode platform and get its correctness and runtime (if accepted). Acceptance of any of the 3 codes is considered as LLM acceptance and the runtime of the first accepted code is recorded, otherwise it is considered as failure. We repeat the submission of each piece of code for 3 times and record the average results.

\subsubsection{Models}
\label{section:exp_rq1_models}

We select commercial and open source LLMs that achieve the SOTA performance on HumanEval and MBPP:

\textbf{GPT-3.5 and GPT-4}. OpenAI's GPT-3.5 and GPT-4 can be seen as the most powerful LLM. We use two models by using the OpenAI API with model ids \texttt{gpt-3.5-turbo-1106} and \texttt{gpt-4-1106-preview}.

\textbf{Phi-2}. Phi-2~\cite{phi-2} is a 2.7B-parameter model that demonstrates outstanding reasoning and language understanding capabilities, showcasing SOTA performance among LLMs smaller than 13B.

\textbf{Code Llama}. Code Llama~\cite{rozière2023codellama} is built on top of Llama 2~\cite{touvron2023llama} and is fine-tuned for generating and discussing code. The 7B version is shown to outperform Llama 2 70B on both HumanEval and MBPP.

\textbf{WizardCoder}. WizardCoder~\cite{luo2023wizardcoder} empowers code LLMs with complex instruction fine-tuning and outperforms the largest closed LLMs, Anthropic's Claude~\cite{claude} and Google's Bard~\cite{bard}, on HumanEval.

\textbf{DeepSeek Coder}. DeepSeek Coder~\cite{bi2024deepseek,deepseek-coder} 33B version is able to outperform GPT-3.5 on HumanEval and achieve comparable results with GPT-3.5 on MBPP after instruct tuning. We choose the 33B version of DeepSeek Coder before and after instruct tuning for the experiment, denoted as the ``base'' and ``instruct'', respectively.

For LeetCodeEval, since it requires a chat/instruction model, we choose \textbf{GPT-4}, \textbf{GPT-3.5} and \textbf{DeepSeek Coder 33B Instruct}, which perform the best on LeetCode problems in our pre-experiments.

\subsubsection{Metrics}
\label{section:exp_rq1_metrics}

We report average normalized runtime and Pass@10. Pass@10 metric is the probability that at least one of the top 10-generated code samples for a problem passes all test cases.

Since there is no runtime on failed problem, we compute runtime metric only for problems where all LLMs pass. For each such problem, we count the runtime of each code on all test cases. Let $t(M_j)^i=\{t(M_j)^i_1,...,t(M_j)^i_n\}$ denotes the runtime of the code $c^j$ generated by LLM $M_j$ on test cases $tc^i=\{tc^i_1,...tc^i_n\}$, where $n$ is the number of test cases of problem $p^i$. Then, following the practice of online programming websites such as LeetCode~\cite{leetcode} and Codeforces~\cite{codeforces}, we take the longest of these runtimes, i.e., $max(t^i)$, as the final runtime of the LLM $M_j$ on the problem $p^i$.

Nevertheless, there are order of magnitude differences in the runtime of LLM on different problems, we normalize all the runtimes of all LLMs on each problem. Concretely, for the problem $p^i$, we let $t(M)^i=\{t(M_1)^i,...,t(M_l)^i\}$ be the runtime of LLMs $M=\{m_1,...,m_l\}$ on problem $p^i$, where $l$ is the number of LLMs. Then the normalized runtime of LLMs on problem $p^i$ is calculated as $normalize(t(M)^i)=\{\frac{t(M_1)^i}{sum(t(M)^i)},...,\frac{t(M_l)^i}{sum(t(M)^i)}\}$, where $sum(t(M)^i)$ denotes the summarization of all elements in $t(M)^i$. Then, the average normalized runtime of the LLM $M_j$ is denoted as $\frac{\sum_{i=1}^{o}{normalize(t(M_j)^i)}}{o}$, where $o$ is the number of problems that all LLMs pass, $normalize(t(M_j)^i)$ is the normalized runtime of the LLM $M_j$ on the problem $p^i$.

Besides, we also adopt average percentage beats for LeetCodeEval, i.e., the average of the percentage of each accepted code that beats the other users.

\subsubsection{Results}
\label{section:exp_rq1_results}

\begin{table}[t!]
    \centering
    \caption{Results on HumanEval and MBPP.}
    \label{table:rq1_results}
    \resizebox{\linewidth}{!}{%
        \begin{tabular}{llrrrr}
            \toprule
                \multirow{2}{*}{LLM} &
                \multirow{2}{*}{Version} &
                \multicolumn{2}{c}{HumanEval} &
                \multicolumn{2}{c}{MBPP} \\
            \cline{3-6}
                &
                &
                \multicolumn{1}{c}{Runtime} &
                \multicolumn{1}{c}{Pass@10} &
                \multicolumn{1}{c}{Runtime} &
                \multicolumn{1}{c}{Pass@10} \\
            \midrule
                GPT-4                        & N/A          & 8.61 & 98.2 & 9.14 & 94.2 \\
                GPT-3.5                      & N/A          & 8.35 & 87.2 & 8.86 & 88.7 \\
            \hline
                Phi-2                        & 2.7B         & 8.78 & 62.8 & 8.98 & 74.7 \\
            \hline
                \multirow{3}{*}{Code Llama}  & 7B           & 9.95 & 68.9 & 9.58 & 81.0 \\
                                             & 13B          & 9.87 & 79.3 & 9.61 & 83.0 \\
                                             & 34B          & 9.93 & 80.5 & 9.54 & 85.0 \\
            \hline
                \multirow{3}{*}{WizardCoder} & 7B           & 9.35 & 67.7 & 8.54 & 74.4 \\
                                             & 13B          & 9.18 & 75.0 & 8.83 & 81.5 \\
                                             & 34B          & 9.04 & 83.5 & 8.60 & 85.5 \\
            \hline
                \multirow{2}{*}{DeepSeek Coder} & 33B Base  & 9.40 & 79.9 & 9.42 & 82.0 \\
                                             & 33B Instruct & 7.54 & 93.9 & 8.93 & 90.0 \\
            \bottomrule
        \end{tabular}%
    }
\end{table}
\begin{table}[htbp]
    \centering
    \small
    \caption{Results of LLMs on easy and medium subsets. GPT-3.5 is excluded on the Medium subset.}
    \label{table:rq2_results}
        \begin{tabular}{lrrrr}
            \toprule
                \multirow{2}{*}{LLM} & \multicolumn{2}{c}{Easy} & \multicolumn{2}{c}{Medium} \\
            \cline{2-5}
                                     & Runtime     & \%Beats    & Runtime      & \%Beats     \\
            \midrule
                GPT-4                & 30.89       & 65.51      & 50.92        & 73.09       \\
                GPT-3.5              & 33.80       & 62.08      & -            & -           \\
                DeepSeek Coder       & 35.30       & 61.05      & 49.08        & 67.46       \\
            \bottomrule
        \end{tabular}%
\end{table}

Table~\ref{table:rq1_results} and Table~\ref{table:rq2_results} shows the results of LLM on HumanEval and MBPP, and LeetCodeEval, respectively. Note that the average normalized runtime is computed based only on the programming problems that all LLMs pass, and there are 70 and 242 problems that all LLMs pass in HumanEval and MBPP, 24, 3 and 0 on easy, medium and hard subsets of LeetCodeEval, respectively. So, we cannot compare models' performances on the hard subset, and for medium subset, we find that 12 medium problems passed by both GPT-4 and DeepSeek Coder, so we only compare and report the two models on medium set.

\textbf{First, the ability to generate correct code is not positively correlated with the ability to generate efficient code.} For example, the Pass@10 of GPT-4 has a clear advantage over GPT-3.5, but the code generated by the former is not as efficient as the latter on both HumanEval and MBPP. The same happens with Phi-2, which, despite having the lowest Pass@10, generates code with a lower runtime than most of the other models. 
\textbf{Second, larger number of parameters does not promise higher performance.} Code Llama and WizardCoder series demonstrate that increasing the number of parameters does not significantly affect the runtime of generated code across models of different sizes. This suggests that models of varying sizes share similar performance due to their reliance on the same training data.
\textbf{Then, training strategy and data have an impact on the efficiency of the generated code.} For example, DeepSeek Coder 33B Instruct has a significant advantage over its Base version. Indeed the ``Base'' version is trained on code corpus by completion and fill-in-the-blank tasks, while the ``Instruct'' version is the result of further instruct-tuning of the ``Base'' version on the instruction data.
\textbf{Last, LLM performs differently across benchmarks.} On HumanEval, DeepSeek Coder 33B Instruct has the lowest runtime, but on MBPP, the lowest model becomes the WizardCoder series. We argue that this is related to the data distribution of the model and the dataset.
In addition, on LeetCodeEval, the code generated by GPT-4 has the highest efficiency on average. We believe this is due to more diverse test cases compared to HumanEval and MBPP. Comprehensive test cases on LeetCode can make the runtime benefits of code with real less complexity more significant, and thus more accurately reflect the efficiency. 

\subsection{RQ2: Prompting for More Efficient Code}
\label{section:exp_rq3}

\begin{figure}
    \centering

\begin{tcolorbox}[title=Prompt 1,fonttitle=\small,fontupper=\footnotesize\ttfamily,left=1mm,right=1mm,top=1mm,bottom=1mm]
\textbf{User}: \{original\_prompt\}\\
Please make the code as time efficient as possible.\\
\textbf{LLM}: \{fast\_code\}
\end{tcolorbox}
\vspace{-1.0em}
\begin{tcolorbox}[title=Prompt 2,fonttitle=\small,fontupper=\footnotesize\ttfamily,left=1mm,right=1mm,top=1mm,bottom=1mm]
\textbf{User}: \{original\_prompt\}\\
\textbf{LLM}: \{slow\_code\}\\
\textbf{User}: Optimize the code and provide a more efficient version.\\
\textbf{LLM}: \{fast\_code\}
\end{tcolorbox}
\vspace{-1.0em}
\begin{tcolorbox}[title=Prompt 3,fonttitle=\small,fontupper=\footnotesize\ttfamily,left=1mm,right=1mm,top=1mm,bottom=1mm]
\textbf{User}: \{original\_prompt\}\\
\textbf{LLM}: \{slow\_code\}\\
\textbf{User}: Give a potential strategy improving the efficiency of the code.\\
\textbf{LLM}: {strategy}\\
\textbf{User}: Now give the optimized version of the same code with the strategy mentioned above.\\
\textbf{LLM}: \{fast\_code\}
\end{tcolorbox}

\vspace{-1.5em}
    \caption{Three prompt methods.}
    \label{figure:rq3_prompts}
\end{figure}

We try three different prompts which are illustrated in Figure~\ref{figure:rq3_prompts}, where the last two prompts are introduced by Madaan et al.~\cite{madaan2023learning}. Prompt 1 directly asks the LLM to generate the code as efficient as possible. Both prompt 2 and prompt 3 are chain-of-thought prompts. They first use the original prompt to make the model generate the original code. Then prompt 2 asks model to optimize it, while prompt 3 first has the model analyze optimization strategies before generating the optimized code.

We apply the three prompts on GPT-4, GPT-3.5 and DeepSeek Coder 33B Instruct. Note that here, same as in 
RQ2, 
both the LeetCodeEval hard subset and the GPT-3.5 on the LeetCodeEval medium subset are excluded from evaluation. For metrics, we choose the speedup rate, i.e., let $t_o$ and $t_n$ be the runtime of the original code and optimized code, respectively, then $speedup=\frac{t_o}{t_n}$. Following Madaan et al.~\cite{madaan2023learning}, for cases where the optimized code fails or the runtime is higher, we make $speedup=1$.

\begin{table}[t!]
    \centering
    \footnotesize
    \caption{Speedup of three prompt methods.}
    \label{table:rq3_results}
        \begin{tabular}{llrrrr}
            \toprule
                \multirow{2}{*}{Method} & \multirow{2}{*}{LLM} & \multirow{2}{*}{HumanEval} & \multirow{2}{*}{MBPP} & \multicolumn{2}{c}{LeetCodeEval} \\
            \cline{5-6}
                                          &                &      &      & Easy & Medium \\
            \midrule
                \multirow{3}{*}{Prompt 1} & GPT-4          & 1.06 & 1.04 & 1.13 & 1.07   \\
                                          & GPT-3.5        & 1.04 & 1.03 & 1.11 & -      \\
                                          & DeepSeek Coder & 1.00 & 1.01 & 1.03 & 1.02   \\
            \hline
                \multirow{3}{*}{Prompt 2} & GPT-4          & 1.06 & 1.05 & 1.15 & 1.16   \\
                                          & GPT-3.5        & 1.03 & 1.03 & 1.15 & -      \\
                                          & DeepSeek Coder & 1.01 & 1.02 & 1.05 & 1.02   \\
            \hline
                \multirow{3}{*}{Prompt 3} & GPT-4          & 1.05 & 1.04 & 1.18 & 1.18   \\
                                          & GPT-3.5        & 1.04 & 1.03 & 1.16 & -      \\
                                          & DeepSeek Coder & 1.01 & 1.00 & 1.05 & 1.01   \\
            \bottomrule
        \end{tabular}%
\end{table}

The overall results are shown in Table~\ref{table:rq3_results}. \textbf{First, the prompt method generally works better on LeetCodeEval than on HumanEval and MBPP.} We believe there are two reasons: (1) HumanEval and MBPP problems have lower average difficulty and complexity than LeetCodeEval, resulting in a constrained optimization space and similar performance across prompt methods, and (2) the limited input size of the former prevents the reduction in algorithmic complexity from being evident in the runtime, whereas the more extensive test cases in LeetCodeEval magnify the performance of code with lower complexity. \textbf{Second, the three prompts have a larger gap on the medium subset of LeetCodeEval than easy subset.} This is because the simplicity of the easy subset allows the model to produce correct and efficient code simultaneously. However, the increased complexity in the medium subset, due to higher problem difficulty, hinders Prompt 1 from generating compliant code in a single step. Improved results are achieved by having the model first generate correct code and then analyze and optimize it step by step.

\section{Threats to Validity}
\label{section:threats}

Potential data leakage is a threat to construct validity because we can not know if the data used for evaluation is present in the training data of models. We mitigate this threat by selecting only LeetCode problems after April 2023, which is the latest knowledge cut-off of the GPT series, however, we are unable to get the data cut-offs for the other model. Threats to internal validity is related to the unstable runtime, we mitigate this by using the gem5 CPU simulator and running each evaluation process multiple times.

\section{Related Work}
\label{section:related}



DeepDev-PERF~\cite{garg2022deepdev}, a deep learning-based approach to
improve software performance for C\# applications, can generate the same performance improvement suggestions as the developer patches in 53\% of the cases. Madaan et al.~\cite{madaan2023learning} adapt LLMs to code optimization with respect to the runtime. They propose PIE, a dataset consists of C++ program pairs with runtime annotations, and evaluate different prompting and fine-tuning approaches for adapting LLMs to optimize programs. By allowing LLMs to iteratively provide self-feedback and refine their own outputs, Self-Refine~\cite{madaan2023self} increases the LLM's performace on PIE dataset.

Rather than efficiency, Siddiq et al.~\cite{siddiq2023lightweight} evaluate and improve the quality of the automatically generated code by LLMs w.r.t. the adherence to coding standards and presence of code smells and security smells. Yetiştiren et al.~\cite{yeticstiren2023evaluating} assesses the code generation capabilities of several LLMs in terms of code quality metrics, such as code validity and maintainability.

\section{Conclusion and Future Work}
\label{section:conclusion}

This paper evaluates the efficiency of LLM-generated code, revealing that (1) the efficiency of LLM-generated code is independent of the model's performance on generating correct code and model size, and (2) step-by-step prompting could make LLM to generate more efficient code, especially on complex problems. Our study suggests a research avenue for improving LLMs in code efficiency, offering practical insights for model selection. Future work will focus on proposing a novel prompt method to enhance LLM-generated code efficiency.

\begin{acks}

This research is supported by the Cooperation Fund of Huawei-NJU Creative Laboratory for the Next Programming, CCF-Huawei Populus Grove Fund, NSF award 2034508. We also thank the reviewers for their helpful comments. Chuanyi Li is the corresponding author.
\end{acks}

\bibliographystyle{ACM-Reference-Format}
\bibliography{refs}


\begin{thebibliography}{27}


\ifx \showCODEN    \undefined \def \showCODEN     #1{\unskip}     \fi
\ifx \showDOI      \undefined \def \showDOI       #1{#1}\fi
\ifx \showISBNx    \undefined \def \showISBNx     #1{\unskip}     \fi
\ifx \showISBNxiii \undefined \def \showISBNxiii  #1{\unskip}     \fi
\ifx \showISSN     \undefined \def \showISSN      #1{\unskip}     \fi
\ifx \showLCCN     \undefined \def \showLCCN      #1{\unskip}     \fi
\ifx \shownote     \undefined \def \shownote      #1{#1}          \fi
\ifx \showarticletitle \undefined \def \showarticletitle #1{#1}   \fi
\ifx \showURL      \undefined \def \showURL       {\relax}        \fi
\providecommand\bibfield[2]{#2}
\providecommand\bibinfo[2]{#2}
\providecommand\natexlab[1]{#1}
\providecommand\showeprint[2][]{arXiv:#2}

\bibitem[eff({[n.\,d.]})]%
        {efficiencyeval}
 \bibinfo{year}{[n.\,d.]}\natexlab{}.
\newblock \bibinfo{howpublished}{\url{https://github.com/NougatCA/EfficiencyEval}}.
\newblock


\bibitem[cod({[n.\,d.]})]%
        {codeforces}
 \bibinfo{year}{[n.\,d.]}\natexlab{}.
\newblock \bibinfo{title}{Codeforces}.
\newblock \bibinfo{howpublished}{\url{https://codeforces.com/}}.
\newblock


\bibitem[lee({[n.\,d.]})]%
        {leetcode}
 \bibinfo{year}{[n.\,d.]}\natexlab{}.
\newblock \bibinfo{title}{LeetCode}.
\newblock \bibinfo{howpublished}{\url{https://leetcode.com/}}.
\newblock


\bibitem[Akram and Sawalha(2019)]%
        {akram2019validation}
\bibfield{author}{\bibinfo{person}{Ayaz Akram} {and} \bibinfo{person}{Lina Sawalha}.} \bibinfo{year}{2019}\natexlab{}.
\newblock \showarticletitle{Validation of the gem5 simulator for x86 architectures}. In \bibinfo{booktitle}{\emph{2019 IEEE/ACM Performance Modeling, Benchmarking and Simulation of High Performance Computer Systems (PMBS)}}. IEEE, \bibinfo{pages}{53--58}.
\newblock


\bibitem[Anthropic({[n.\,d.]})]%
        {claude}
\bibfield{author}{\bibinfo{person}{Anthropic}.} \bibinfo{year}{[n.\,d.]}\natexlab{}.
\newblock \bibinfo{title}{Introducing Claude}.
\newblock \bibinfo{howpublished}{\url{https://www.anthropic.com/index/introducing-claude}}.
\newblock


\bibitem[Austin et~al\mbox{.}(2021)]%
        {austin2021mbpp}
\bibfield{author}{\bibinfo{person}{Jacob Austin}, \bibinfo{person}{Augustus Odena}, \bibinfo{person}{Maxwell Nye}, \bibinfo{person}{Maarten Bosma}, \bibinfo{person}{Henryk Michalewski}, \bibinfo{person}{David Dohan}, \bibinfo{person}{Ellen Jiang}, \bibinfo{person}{Carrie Cai}, \bibinfo{person}{Michael Terry}, \bibinfo{person}{Quoc Le}, {and} \bibinfo{person}{Charles Sutton}.} \bibinfo{year}{2021}\natexlab{}.
\newblock \bibinfo{title}{Program Synthesis with Large Language Models}.
\newblock
\newblock
\showeprint[arxiv]{2108.07732}~[cs.PL]


\bibitem[Bi et~al\mbox{.}(2024)]%
        {bi2024deepseek}
\bibfield{author}{\bibinfo{person}{Xiao Bi}, \bibinfo{person}{Deli Chen}, \bibinfo{person}{Guanting Chen}, \bibinfo{person}{Shanhuang Chen}, \bibinfo{person}{Damai Dai}, \bibinfo{person}{Chengqi Deng}, \bibinfo{person}{Honghui Ding}, \bibinfo{person}{Kai Dong}, \bibinfo{person}{Qiushi Du}, \bibinfo{person}{Zhe Fu}, {et~al\mbox{.}}} \bibinfo{year}{2024}\natexlab{}.
\newblock \showarticletitle{DeepSeek LLM: Scaling Open-Source Language Models with Longtermism}.
\newblock \bibinfo{journal}{\emph{arXiv preprint arXiv:2401.02954}} (\bibinfo{year}{2024}).
\newblock


\bibitem[Binkert et~al\mbox{.}(2011)]%
        {binkert2011gem5}
\bibfield{author}{\bibinfo{person}{Nathan Binkert}, \bibinfo{person}{Bradford Beckmann}, \bibinfo{person}{Gabriel Black}, \bibinfo{person}{Steven~K Reinhardt}, \bibinfo{person}{Ali Saidi}, \bibinfo{person}{Arkaprava Basu}, \bibinfo{person}{Joel Hestness}, \bibinfo{person}{Derek~R Hower}, \bibinfo{person}{Tushar Krishna}, \bibinfo{person}{Somayeh Sardashti}, {et~al\mbox{.}}} \bibinfo{year}{2011}\natexlab{}.
\newblock \showarticletitle{The gem5 simulator}.
\newblock \bibinfo{journal}{\emph{ACM SIGARCH computer architecture news}} \bibinfo{volume}{39}, \bibinfo{number}{2} (\bibinfo{year}{2011}), \bibinfo{pages}{1--7}.
\newblock


\bibitem[Chen et~al\mbox{.}(2023)]%
        {chen2023codet}
\bibfield{author}{\bibinfo{person}{Bei Chen}, \bibinfo{person}{Fengji Zhang}, \bibinfo{person}{Anh Nguyen}, \bibinfo{person}{Daoguang Zan}, \bibinfo{person}{Zeqi Lin}, \bibinfo{person}{Jian-Guang Lou}, {and} \bibinfo{person}{Weizhu Chen}.} \bibinfo{year}{2023}\natexlab{}.
\newblock \showarticletitle{CodeT: Code Generation with Generated Tests}. In \bibinfo{booktitle}{\emph{The Eleventh International Conference on Learning Representations}}.
\newblock
\urldef\tempurl%
\url{https://openreview.net/forum?id=ktrw68Cmu9c}
\showURL{%
\tempurl}


\bibitem[Chen et~al\mbox{.}(2021)]%
        {chen2021humaneval}
\bibfield{author}{\bibinfo{person}{Mark Chen}, \bibinfo{person}{Jerry Tworek}, \bibinfo{person}{Heewoo Jun}, \bibinfo{person}{Qiming Yuan}, \bibinfo{person}{Henrique Ponde de~Oliveira Pinto}, \bibinfo{person}{Jared Kaplan}, \bibinfo{person}{Harri Edwards}, \bibinfo{person}{Yuri Burda}, \bibinfo{person}{Nicholas Joseph}, \bibinfo{person}{Greg Brockman}, {et~al\mbox{.}}} \bibinfo{year}{2021}\natexlab{}.
\newblock \bibinfo{title}{Evaluating Large Language Models Trained on Code}.
\newblock
\newblock
\showeprint[arxiv]{2107.03374}~[cs.LG]


\bibitem[DeepSeek(2023)]%
        {deepseek-coder}
\bibfield{author}{\bibinfo{person}{DeepSeek}.} \bibinfo{year}{2023}\natexlab{}.
\newblock \bibinfo{title}{DeepSeek Coder: Let the Code Write Itself}.
\newblock \bibinfo{howpublished}{\url{https://github.com/deepseek-ai/DeepSeek-Coder}}.
\newblock


\bibitem[Garg et~al\mbox{.}(2022)]%
        {garg2022deepdev}
\bibfield{author}{\bibinfo{person}{Spandan Garg}, \bibinfo{person}{Roshanak~Zilouchian Moghaddam}, \bibinfo{person}{Colin~B Clement}, \bibinfo{person}{Neel Sundaresan}, {and} \bibinfo{person}{Chen Wu}.} \bibinfo{year}{2022}\natexlab{}.
\newblock \showarticletitle{DeepDev-PERF: a deep learning-based approach for improving software performance}. In \bibinfo{booktitle}{\emph{Proceedings of the 30th ACM Joint European Software Engineering Conference and Symposium on the Foundations of Software Engineering}}. \bibinfo{pages}{948--958}.
\newblock


\bibitem[GitHub({[n.\,d.]})]%
        {copilot}
\bibfield{author}{\bibinfo{person}{GitHub}.} \bibinfo{year}{[n.\,d.]}\natexlab{}.
\newblock \bibinfo{title}{GitHub Copilot}.
\newblock \bibinfo{howpublished}{\url{https://github.com/features/copilot}}.
\newblock


\bibitem[Google({[n.\,d.]})]%
        {bard}
\bibfield{author}{\bibinfo{person}{Google}.} \bibinfo{year}{[n.\,d.]}\natexlab{}.
\newblock \bibinfo{title}{Bard}.
\newblock \bibinfo{howpublished}{\url{https://bard.google.com/}}.
\newblock


\bibitem[Huang et~al\mbox{.}(2023)]%
        {huang2023anpl}
\bibfield{author}{\bibinfo{person}{Di Huang}, \bibinfo{person}{Ziyuan Nan}, \bibinfo{person}{Xing Hu}, \bibinfo{person}{Pengwei Jin}, \bibinfo{person}{Shaohui Peng}, \bibinfo{person}{Yuanbo Wen}, \bibinfo{person}{Rui Zhang}, \bibinfo{person}{Zidong Du}, \bibinfo{person}{Qi Guo}, \bibinfo{person}{Yewen Pu}, {and} \bibinfo{person}{Yunji Chen}.} \bibinfo{year}{2023}\natexlab{}.
\newblock \showarticletitle{{ANPL}: Towards Natural Programming with Interactive Decomposition}. In \bibinfo{booktitle}{\emph{Thirty-seventh Conference on Neural Information Processing Systems}}.
\newblock
\urldef\tempurl%
\url{https://openreview.net/forum?id=RTRS3ZTsSj}
\showURL{%
\tempurl}


\bibitem[JetBrains({[n.\,d.]})]%
        {jetbrains}
\bibfield{author}{\bibinfo{person}{JetBrains}.} \bibinfo{year}{[n.\,d.]}\natexlab{}.
\newblock \bibinfo{title}{JetBrains AI}.
\newblock \bibinfo{howpublished}{\url{https://www.jetbrains.com/ai/}}.
\newblock


\bibitem[Liu et~al\mbox{.}(2023)]%
        {liu2023evalplus}
\bibfield{author}{\bibinfo{person}{Jiawei Liu}, \bibinfo{person}{Chunqiu~Steven Xia}, \bibinfo{person}{Yuyao Wang}, {and} \bibinfo{person}{Lingming Zhang}.} \bibinfo{year}{2023}\natexlab{}.
\newblock \showarticletitle{Is Your Code Generated by Chat{GPT} Really Correct? Rigorous Evaluation of Large Language Models for Code Generation}. In \bibinfo{booktitle}{\emph{Thirty-seventh Conference on Neural Information Processing Systems}}.
\newblock


\bibitem[Luo et~al\mbox{.}(2023)]%
        {luo2023wizardcoder}
\bibfield{author}{\bibinfo{person}{Ziyang Luo}, \bibinfo{person}{Can Xu}, \bibinfo{person}{Pu Zhao}, \bibinfo{person}{Qingfeng Sun}, \bibinfo{person}{Xiubo Geng}, \bibinfo{person}{Wenxiang Hu}, \bibinfo{person}{Chongyang Tao}, \bibinfo{person}{Jing Ma}, \bibinfo{person}{Qingwei Lin}, {and} \bibinfo{person}{Daxin Jiang}.} \bibinfo{year}{2023}\natexlab{}.
\newblock \bibinfo{title}{WizardCoder: Empowering Code Large Language Models with Evol-Instruct}.
\newblock
\newblock
\showeprint[arxiv]{2306.08568}~[cs.CL]


\bibitem[Madaan et~al\mbox{.}(2024)]%
        {madaan2023learning}
\bibfield{author}{\bibinfo{person}{Aman Madaan}, \bibinfo{person}{Alexander Shypula}, \bibinfo{person}{Uri Alon}, \bibinfo{person}{Milad Hashemi}, \bibinfo{person}{Parthasarathy Ranganathan}, \bibinfo{person}{Yiming Yang}, \bibinfo{person}{Graham Neubig}, {and} \bibinfo{person}{Amir Yazdanbakhsh}.} \bibinfo{year}{2024}\natexlab{}.
\newblock \showarticletitle{Learning Performance-Improving Code Edits}. In \bibinfo{booktitle}{\emph{International Conference on Learning Representations}}.
\newblock


\bibitem[Madaan et~al\mbox{.}(2023)]%
        {madaan2023self}
\bibfield{author}{\bibinfo{person}{Aman Madaan}, \bibinfo{person}{Niket Tandon}, \bibinfo{person}{Prakhar Gupta}, \bibinfo{person}{Skyler Hallinan}, \bibinfo{person}{Luyu Gao}, \bibinfo{person}{Sarah Wiegreffe}, \bibinfo{person}{Uri Alon}, \bibinfo{person}{Nouha Dziri}, \bibinfo{person}{Shrimai Prabhumoye}, \bibinfo{person}{Yiming Yang}, {et~al\mbox{.}}} \bibinfo{year}{2023}\natexlab{}.
\newblock \showarticletitle{Self-refine: Iterative refinement with self-feedback}.
\newblock \bibinfo{journal}{\emph{arXiv preprint arXiv:2303.17651}} (\bibinfo{year}{2023}).
\newblock


\bibitem[Microsoft(2023)]%
        {phi-2}
\bibfield{author}{\bibinfo{person}{Microsoft}.} \bibinfo{year}{2023}\natexlab{}.
\newblock \bibinfo{title}{Phi-2: The surprising power of small language models}.
\newblock \bibinfo{howpublished}{\url{https://www.microsoft.com/en-us/research/blog/phi-2-the-surprising-power-of-small-language-models/}}.
\newblock


\bibitem[OpenAI(2023)]%
        {openai2023gpt4}
\bibfield{author}{\bibinfo{person}{OpenAI}.} \bibinfo{year}{2023}\natexlab{}.
\newblock \bibinfo{title}{GPT-4 Technical Report}.
\newblock
\newblock
\showeprint[arxiv]{2303.08774}~[cs.CL]


\bibitem[Roziere et~al\mbox{.}(2023)]%
        {rozière2023codellama}
\bibfield{author}{\bibinfo{person}{Baptiste Roziere}, \bibinfo{person}{Jonas Gehring}, \bibinfo{person}{Fabian Gloeckle}, \bibinfo{person}{Sten Sootla}, \bibinfo{person}{Itai Gat}, \bibinfo{person}{Xiaoqing~Ellen Tan}, \bibinfo{person}{Yossi Adi}, \bibinfo{person}{Jingyu Liu}, \bibinfo{person}{Tal Remez}, \bibinfo{person}{J{\'e}r{\'e}my Rapin}, {et~al\mbox{.}}} \bibinfo{year}{2023}\natexlab{}.
\newblock \bibinfo{title}{Code Llama: Open Foundation Models for Code}.
\newblock
\newblock
\showeprint[arxiv]{2308.12950}~[cs.CL]


\bibitem[Siddiq et~al\mbox{.}(2023)]%
        {siddiq2023lightweight}
\bibfield{author}{\bibinfo{person}{Mohammed~Latif Siddiq}, \bibinfo{person}{Beatrice Casey}, {and} \bibinfo{person}{Joanna Santos}.} \bibinfo{year}{2023}\natexlab{}.
\newblock \showarticletitle{A Lightweight Framework for High-Quality Code Generation}.
\newblock \bibinfo{journal}{\emph{arXiv preprint arXiv:2307.08220}} (\bibinfo{year}{2023}).
\newblock


\bibitem[Touvron et~al\mbox{.}(2023)]%
        {touvron2023llama}
\bibfield{author}{\bibinfo{person}{Hugo Touvron}, \bibinfo{person}{Louis Martin}, \bibinfo{person}{Kevin Stone}, \bibinfo{person}{Peter Albert}, \bibinfo{person}{Amjad Almahairi}, \bibinfo{person}{Yasmine Babaei}, \bibinfo{person}{Nikolay Bashlykov}, \bibinfo{person}{Soumya Batra}, \bibinfo{person}{Prajjwal Bhargava}, \bibinfo{person}{Shruti Bhosale}, {et~al\mbox{.}}} \bibinfo{year}{2023}\natexlab{}.
\newblock \showarticletitle{Llama 2: Open foundation and fine-tuned chat models}.
\newblock \bibinfo{journal}{\emph{arXiv preprint arXiv:2307.09288}} (\bibinfo{year}{2023}).
\newblock


\bibitem[Yeti{\c{s}}tiren et~al\mbox{.}(2023)]%
        {yeticstiren2023evaluating}
\bibfield{author}{\bibinfo{person}{Burak Yeti{\c{s}}tiren}, \bibinfo{person}{I{\c{s}}{\i}k {\"O}zsoy}, \bibinfo{person}{Miray Ayerdem}, {and} \bibinfo{person}{Eray T{\"u}z{\"u}n}.} \bibinfo{year}{2023}\natexlab{}.
\newblock \showarticletitle{Evaluating the code quality of ai-assisted code generation tools: An empirical study on github copilot, amazon codewhisperer, and chatgpt}.
\newblock \bibinfo{journal}{\emph{arXiv preprint arXiv:2304.10778}} (\bibinfo{year}{2023}).
\newblock


\bibitem[Zelikman et~al\mbox{.}(2023)]%
        {zelikman2023parsel}
\bibfield{author}{\bibinfo{person}{Eric Zelikman}, \bibinfo{person}{Qian Huang}, \bibinfo{person}{Gabriel Poesia}, \bibinfo{person}{Noah Goodman}, {and} \bibinfo{person}{Nick Haber}.} \bibinfo{year}{2023}\natexlab{}.
\newblock \showarticletitle{Parsel: Algorithmic Reasoning with Language Models by Composing Decompositions}. In \bibinfo{booktitle}{\emph{Thirty-seventh Conference on Neural Information Processing Systems}}.
\newblock
\urldef\tempurl%
\url{https://openreview.net/forum?id=qd9qcbVAwQ}
\showURL{%
\tempurl}


\end{thebibliography}


\end{document}